\pdfoutput=1
\documentclass[prl,twocolumn,showpacs,amsmath,amssymb,floatfix]{revtex4-1}

\usepackage{graphicx}  
\usepackage{dcolumn}   
\usepackage{bm}        
\usepackage{amssymb}   
\usepackage{amsmath}
\usepackage{mathrsfs}
\usepackage{epigraph}
\usepackage{braket}
\usepackage{gensymb}
\usepackage{lmodern}
\usepackage{lipsum}
\usepackage{marvosym}
\usepackage{ tipa }
\usepackage{bbold}
\usepackage{dsfont}
\usepackage{esint}
\usepackage{mathdots}
\usepackage{xcolor}
\usepackage{appendix}
\usepackage{hyperref}
\usepackage{natbib}
\usepackage{multirow}

\DeclareMathOperator{\deff}{\equiv}

\begin{document}
	
\title{Effective spin-mixing conductance of topological-insulator/ferromagnet and heavy-metal/ferromagnet spin-orbit-coupled interfaces: A first-principles Floquet-nonequilibrium-Green-function approach}

\author{Kapildeb Dolui}
\affiliation{Department of Physics and Astronomy, University of Delaware, Newark, DE 19716, USA}
\author{Utkarsh Bajpai}
\affiliation{Department of Physics and Astronomy, University of Delaware, Newark, DE 19716, USA}
\author{Branislav K. Nikoli\'{c}}
\email{bnikolic@udel.edu}
\affiliation{Department of Physics and Astronomy, University of Delaware, Newark, DE 19716, USA}
		
\begin{abstract}
The spin mixing conductance (SMC) is a key quantity determining efficiency of spin transport across interfaces. Thus, knowledge of its precise value is  required for accurate measurement of parameters quantifying numerous effects in spintronics, such as spin-orbit torque, spin Hall magnetoresistance, spin Hall effect and spin pumping. However, the standard expression for SMC, provided by the scattering theory in terms of the reflection probability amplitudes, is inapplicable when strong spin-orbit coupling (SOC) is present directly at the interface. This is the precisely the case of topological-insulator/ferromagnet and heavy-metal/ferromagnet interfaces of great contemporary interest. We introduce an approach where first-principles Hamiltonian of these interfaces, obtained from noncollinear density functional theory (ncDFT) calculations, is combined with charge conserving Floquet-nonequilibrium-Green-function formalism to compute {\em directly} the pumped spin current $I^{S_z}$ into semi-infinite left lead of two-terminal heterostructures Cu/X/Co/Cu or Y/Co/Cu---where X=Bi$_2$Se$_3$ and Y=Pt or W---due to microwave-driven steadily precessing magnetization of the Co layer. This allows us extract an {\em effective SMC} as a prefactor in $I^{S_z}$ vs. precession cone angle $\theta$ dependence, as long as it remains the same, $I^{S_z} \propto \sin^2 \theta$, as in the case where SOC is absent. By comparing calculations where SOC in switched off vs. switched on in ncDFT calculations, we find that SOC consistently {\em reduces} the pumped spin current and, therefore, the effective SMC.
\end{abstract}
\maketitle

{\em Introduction}.---The spin-mixing conductance (SMC) $g_{\uparrow \downarrow}$ is a quantity introduced by the scattering theory~\cite{Brataas2006,Tserkovnyak2005} of quantum transport of electrons to describe behavior of their spin as they reflect at interfaces between a ferromagnetic metal (FM) and a normal metal (NM). It is defined as~\cite{Brataas2006,Tserkovnyak2005}
\begin{equation}\label{eq:smc}
g_{\uparrow \downarrow} = \sum_{n,m} [\delta _{nm} - (r_{nm}^{\uparrow\uparrow})(r_{nm}^{\downarrow\downarrow})^{*}],
\end{equation}
where $r^{\sigma\sigma}_{mn}$ is the reflection probability amplitude for electron in incoming conducting channel $n$ with spin $\sigma = \uparrow , \downarrow$ to end up in outgoing channel $m$ with spin $\sigma$. For example, when an electron impinges from the NM side onto the NM/FM interface, with its spin noncollinear to FM magnetization, the spin will rotate upon reflection with its transverse components (with respect to the magnetization) being given by the real and imaginary part of $g_{\uparrow \downarrow}$~\cite{Brataas2006}. Although $g_{\uparrow \downarrow}$ is a complex number, for Ohmic NM/FM interfaces $\mathrm{Im}(g_{\uparrow \downarrow})$  becomes negligible due to random phases of complex numbers $r^{\sigma\sigma}_{mn}$~\cite{Xia2002,Zwierzycki2005,Carva2007,Zhang2011d}.

Since interfacial spin transport plays a key role in numerous spintronic phenomena---such as spin-transfer~\cite{Brataas2006} and spin-orbit~\cite{Wang2018b,Ramaswamy2018,Manchon2019} torques, spin Seebeck effect~\cite{Adachi2013}, spin Hall magnetoresistance~\cite{Avci2015}, spin pumping~\cite{Tserkovnyak2005} and the inverse spin Hall effect driven by spin pumping~\cite{Saitoh2006,Mosendz2010,Rojas-Sanchez2014}---precise value of $g_{\uparrow \downarrow}$ is crucial to extract correctly  parameters quantifying these effects from experimental measurements~\cite{Zhu2019}. For example, using an overestimated value of $g_{\uparrow \downarrow}$ yields underestimated value of the spin Hall conductivity and the corresponding spin Hall angle. 

\begin{figure}
	\centering
	\includegraphics[scale=0.32,angle=0]{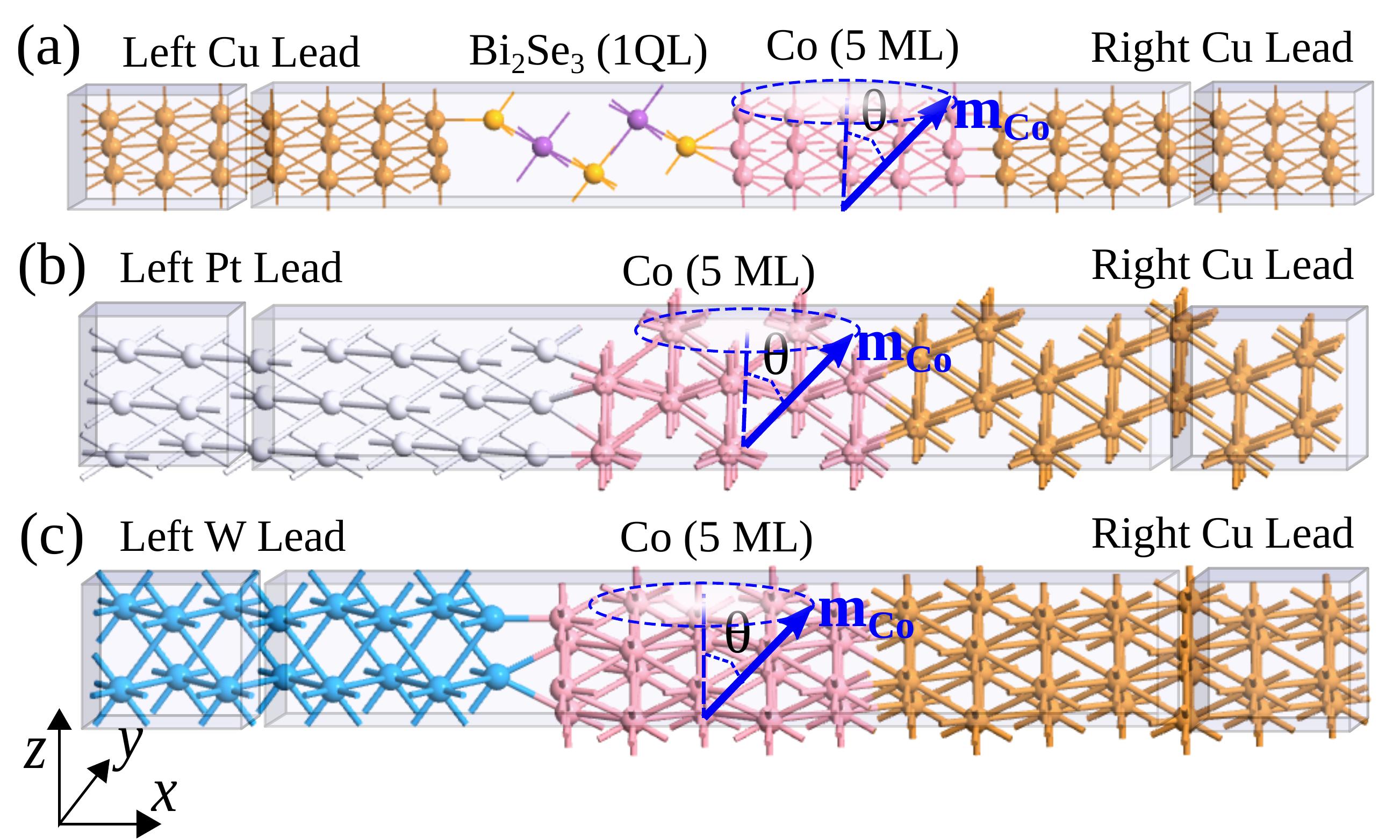}
	\caption{Schematic view of vertical heterostructures whose TI/FM or HM/FM bilayer, with FM composed of 4 monolayers (ML) of Co(0001) with magnetization $\bold{m}_\mathrm{Co}$ precessing with frequency $\omega$ and cone angle $\theta$ around the $z$-axis, is sandwiched between two semi-infinite NM or HM leads: (a) TI=1 quintuple layer (QL) or 6 QL of Bi$_2$Se$_3$(0001) with left and right Cu(111) NM leads; (b) HM = 4 ML of Pt(111) with left Pt(111) lead and right Cu(111) lead; and (c) HM=4 ML of W(001) with left W(001) lead and right Cu(111) lead. The heterostructures are assumed to be infinite in the transverse direction, so that the depicted supercells are periodically repeated within the $yz$-plane. The semi-infinite leads terminate into macroscopic reservoirs without any bias voltage applied between them.}
	\label{fig:fig1}
\end{figure}

In particular, $g_{\uparrow \downarrow}$ governs directly the magnitude of spin current $\mathbf{I}^S=(I^{S_x},I^{S_y},I^{S_z})$~\cite{Tserkovnyak2005}
\begin{equation}\label{eq:ipump}
\mathbf{I}^S = \frac{\hbar} {4\pi} \Big [ \mathrm{Re}(g_{\uparrow\downarrow})~\bold m \times \frac{d\bold m}{dt}  +   \mathrm{Im}(g_{\uparrow\downarrow})~\frac{d\bold m}{dt} \Big ],
\end{equation}
pumped (in the absence of any bias voltage) by precessing magnetization of FM layer into the adjacent NM layer. The precession is typically driven by the absorption of microwaves under the ferromagnetic resonance conditions. Here $\mathbf{m}$ is a unit vector in the direction of the magnetization, viewed as macrospin,  and $\omega$ is the frequency of its precession (as well as of microwaves driving it) around the $z$-axis in the plane of the interface. While $I^{S_x}$ and $I^{S_y}$ oscillate harmonically in time, $I^{S_z}$ is steady or DC spin current.

The spin pumping is interfacial phenomenon in the sense that its magnitude increases with decreasing thickness of the FM layer~\cite{Chen2009}. Since spin pumping means loss of angular momentum (assuming that NM layer is spin sink which prevents backflow of spins into the FM layer~\cite{Tserkovnyak2005}), standard experimental procedure~\cite{Zhu2019} to estimate $g_{\uparrow\downarrow}$ fits Gilbert damping $\alpha$ as a function of the thickness $d_\mathrm{FM}$ of the FM layer by a formula
\begin{equation}\label{eq:alpha}
\alpha=\alpha_\mathrm{intr} + \frac{g_{\uparrow\downarrow}}{d_\mathrm{FM}}\frac{g \mu_B}{4 \pi M_s}.
\end{equation} 
Here $\alpha_\mathrm{intr}$ is the thickness-independent intrinsic damping~\cite{Barati2014,Barati2015} of the FM material, $M_s$ is its saturation magnetization, $g$ is the gyromagnetic ratio, and $\mu_B$ the Bohr magnetron. However, other mechanisms---such as two-magnon scattering~\cite{Zhu2019} and spin memory loss~\cite{Rojas-Sanchez2014,Dolui2017,Belashchenko2016}---can add contributions to $\alpha$, so that using Eq.~\eqref{eq:alpha} to fit experimental data can lead to substantial overestimate of $g_{\uparrow\downarrow}$~\cite{Zhu2019}.

This problem can be resolved by using $g_{\uparrow\downarrow}$ obtained from first-principles quantum transport calculations in a formula for $\alpha$ with two additional terms~\cite{Zhu2019} describing two-magnon scattering and spin memory loss whose values are then fitted to experimental data. However, Eq.~\eqref{eq:ipump} {\em cannot} be used to obtain pumped current from NM/FM interfaces when strong spin-orbit coupling (SOC) is present directly at the interface~\cite{Tserkovnyak2005,Liu2014a} [i.e., interfacial SOC will render $r^{\uparrow \downarrow}_{mn}$ and $r^{\downarrow \uparrow}_{mn}$ elements of the scattering matrix nonzero which is not taken into account by Eqs.~\eqref{eq:smc} and ~\eqref{eq:ipump}]. For example, this is the case of topological-insulator/FM (TI/FM) and heavy-metal/FM (HM/FM) interfaces employed in spin-orbit torque~\cite{Wang2018b,Ramaswamy2018,Manchon2019} and spin-pumping-to-charge conversion~\cite{Deorani2014,Shiomi2014,Jamali2015,Baker2015,Kondou2016,Soumyanarayanan2016} devices where strong SOC is introduced by TI or HM layers, as illustrated in Fig.~\ref{fig:fig1}. For such devices, first-principles quantum transport methods developed to compute $g_{\uparrow \downarrow}$ by combining~\cite{Xia2002,Zwierzycki2005,Zhang2011d} Eq.~\eqref{eq:smc}, or its equivalent~\cite{Carva2007} in terms of steady-state nonequilibrium Green functions (NEGF)~\cite{Stefanucci2013}, with density functional theory (DFT) are inapplicable.  

Nonetheless, an effective SMC, $g_{\uparrow \downarrow}^\mathrm{eff}$, can be extracted as the prefactor of DC pumped spin current vs. precession cone angle $\theta$ [see Fig.~\ref{fig:fig1} for illustration], as long as one can compute $I^{S_z}(\theta)$ directly and this dependence remains the same, $I^{S_z}(\theta) = \frac{\hbar \omega}{4 \pi} g_{\uparrow \downarrow}^\mathrm{eff} \sin^2\theta$, as in Eq.~\eqref{eq:ipump} for interfaces without SOC.  The methods develop to obtain $I^{S_z}(\theta)$ in the presence of interfacial SOC include the Floquet-nonequilibrium Green function (Floquet-NEGF)~\cite{Mahfouzi2012,Mahfouzi2014} and the Kubo formalisms~\cite{Chen2015}. However, they have only been applied to simplistic models of FM/NM bilayers where SOC, such as the Rashba one~\cite{Manchon2015}, is introduced by hand strictly at the interface~\cite{Jamali2015,Mahfouzi2012,Mahfouzi2014,Chen2015}. On the other hand, realistic HM/FM or TI/FM bilayers involve hybridization of wavefunctions of the two materials and proximity effects~\cite{Dolui2017,Marmolejo-Tejada2017} introducing SOC into the FM side or magnetic ordering on the HM and TI side. These effects can hardly be captured by simplistic models and, instead, requires first-principles Hamiltonians.

In this Letter, we introduce a framework  which {\em directly and nonperturbatively} computes pumped spin current across interfaces with strong SOC between realistic materials by combining first-principles Hamiltonian, obtained from noncollinear DFT (ncDFT) calculations~\cite{Capelle2001,Eich2013a}, with charge conserving Floquet-NEGF formalism~\cite{Mahfouzi2012,Mahfouzi2014}. To illustrate applications of such Floquet-NEGF+ncDFT framework, we consider TI/FM and HM/FM interfaces with TI=Bi$_2$Se$_3$ and HM=Pt or W, which are of  great contemporary experimental~\cite{Zhu2019} and technological~\cite{Soumyanarayanan2016} interest.  The two-terminal heterostructures hosting such interfaces are illustrated in Fig.~\ref{fig:fig1}---Cu/Bi$_2$Se$_3$/Co/Cu [Fig.~\ref{fig:fig1}(a)], Pt/Co/Cu [Fig.~\ref{fig:fig1}(b)] and W/Co/Cu [Fig.~\ref{fig:fig1}(c)]---where precessing magnetization of Co FM layer pumps spin current into the semi-infinite Cu or Pt or W leads in the absence of any bias voltage between the macroscopic reservoirs into which such leads terminate. Note that instead of semi-infinite-NM/semi-infinite-FM bilayers device geometry used in the direct computation of SMC~\cite{Xia2002,Zwierzycki2005,Carva2007,Zhang2011d} via Eq.~\eqref{eq:smc} in the absence of SOC, we use two semi-infinite Cu NM leads in Fig.~\ref{fig:fig1}(a) because TI layer is insulating in the bulk and cannot be semi-infinite. Similarly, in order to avoid semi-infinite FM layer, which would  require time-dependent right macroscopic reservoir that is cumbersome to handle in NEGF-based time-dependent quantum transport theory~\cite{Gaury2014}, we insert right semi-infinite Cu lead in Figs.~\ref{fig:fig1}(b) and ~\ref{fig:fig1}(c). 

\begin{figure*}
	\centering
	\includegraphics[scale=0.45]{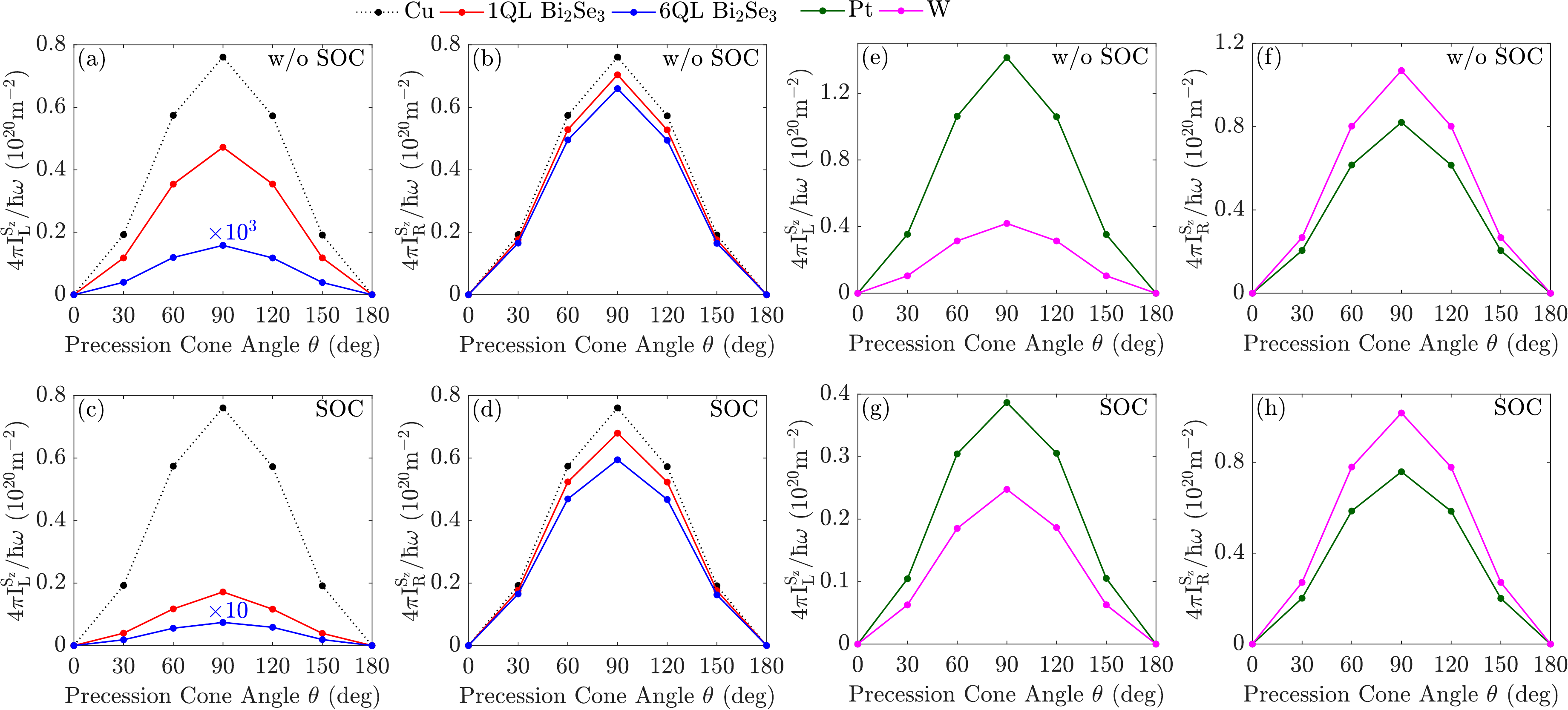}
	\caption{The angular dependence of DC spin currents injected by precessing magnetization of 4 ML of Co into the left, $I_\mathrm{L}^{S_z}$, and the right, $I_\mathrm{R}^{S_z}$, semi-infinite leads  of TI/FM and HM/FM heterostructures in Fig.~\ref{fig:fig1}. The type of TI (such as 1 QL or 6 QL of Bi$_2$Se$_3$) or HM layer (such as 4 ML of Pt or W) is indicated by the legend on the top.  Upper panels (a),(b),(e),(f) and lower panels (c),(d),(g),(h) indicate that SOC is either switched off or switched on in ncDFT calculations, respectively. The dotted black line is for semi-infinite-Cu/Co/semi-infinite-Cu heterostructure~\cite{Carva2007} used as a reference. Note that in the case of 6QL-Bi$_2$Se$_3$/Co heterostructure, the curve is multiplied by a factor 10$^3$ or $10$ in panels (a) and (c), respectively. We use  $k_x \times k_y=25 \times 25$ grid, $N_\mathrm{max}=2$ and  \mbox{$\hbar\omega = 1$ meV} in all panels.}
	\label{fig:fig2}
\end{figure*}

The pumped spin current into the left (L) NM lead of these heterostructures exhibit standard angular dependence~\cite{Tserkovnyak2005} $I^{S_z}_\mathrm{L} \propto \sin^2 \theta$, as confirmed in Fig.~\ref{fig:fig2}. This makes it possible to extract the {\em effective SMC} as
\begin{equation}\label{eq:smceffective}
g^\mathrm{eff}_{\uparrow \downarrow} \deff  \frac{4\pi}{\hbar\omega}I^{S_z}_\mathrm{L} \bigg\rvert_{\theta = 90^\circ}.
\end{equation}
Thus obtained values for $g^\mathrm{eff}_{\uparrow \downarrow}$ are given in Table~\ref{tab:table1}.

{\em First-principles Floquet-Hamiltonian from ncDFT calculations}.---The equilibrium single-particle spin-dependent Kohn-Sham (KS) Hamiltonian of heterostructures in Fig.~\ref{fig:fig1} is given by
\begin{multline}\label{eq:ham_ks}
\bold{H}_\mathrm{KS} = -\frac{\hbar^2\nabla^2}{2m} + \bold{V}_\mathrm{H} + \bold{V}_\mathrm{ext} + \bold{V}_\mathrm{XC}  
+ \bold{V}_\mathrm{SOC} - \boldsymbol{\sigma}\cdot\bold{B}_\mathrm{XC},
\end{multline}
of the non-collinear DFT (ncDFT)~\cite{Capelle2001,Eich2013a}. Here ${\rm \bold{V}}_\mathrm{H}(\bold{r})$, ${\rm \bold{V}}_\mathrm{ext}(\bold{r})$, and ${\rm \bold{V}}_\mathrm{XC}(\bold{r}) = \delta E_\mathrm{XC}[n(\bold{r}),\bold{m}(\bold{r})]/\delta n(\bold{r})$ are Hartree potential, external potential and the exchange-correlation (XC) potential, respectively; $\bold{V}_\mathrm{SOC}$ is additional potential due to SOC; $\boldsymbol{\sigma} = (\hat{\sigma}_x, \hat{\sigma}_y,\hat{\sigma}_z)$ is the vector of the Pauli matrices;  and the XC magnetic field, \mbox{$\mathbf{B}_\mathrm{XC}(\mathbf{r}) = \delta E_\mathrm{XC}[n(\mathbf{r}),\mathbf{m}(\mathbf{r})]/\delta \mathbf{m}(\mathbf{r})$}, is functional derivative with respect to the vector  magnetization density $\mathbf{m}(\mathbf{r})$. The extension of DFT to the case of spin-polarized systems is formally derived in terms of $\mathbf{m}(\mathbf{r})$ and total electron density $n(\mathbf{r})$, where in collinear DFT $\mathbf{m}(\mathbf{r})$ points in the same direction at all points in space, while in ncDFT $\mathbf{m}(\mathbf{r})$ can point in an arbitrary direction~\cite{Capelle2001,Eich2013a}. The matrix representation of the XC magnetic field can be extracted from the $\bold{H}_\mathrm{KS}$ matrix as   \mbox{$\mathbf{B}_\mathrm{XC}=(2 \mathrm{Re}[\mathscr{H}^{\uparrow \downarrow}], -2\mathrm{Im}[\mathscr{H}^{\uparrow\downarrow}], \mathscr{H}^{\uparrow\uparrow} -  \mathscr{H}^{\downarrow \downarrow})$}, where $\mathscr{H} = \bold{H}_\mathrm{KS} - \bold{V}_\mathrm{SOC}$. The time-dependent Hamiltonian of heterostructures in Fig.~\ref{fig:fig1} is then constructed as 
\begin{equation}\label{eq:td_ham}
\bold{H}(t) = \bold{H}_0(\theta) + \bold{V}(\theta)e^{i\omega t} + \bold{V}^\dagger(\theta) e^{-i\omega t},
\end{equation}
where $\bold{H}_0$ and $\bold{V}$ are obtained from \textit{time-independent} ncDFT calculations as $\bold{H}_0(\theta) = \bold{H}_\mathrm{KS}(\theta) - \bold{V}(\theta) - \bold{V}^\dagger(\theta)$ and $\bold{V}(\theta) = \frac{1}{4}\bold{B}_\mathrm{XC}^x(\theta)\otimes [ \hat{\sigma}_x - i\hat{\sigma}_y ]$.

We employ the interface builder in the QuantumATK~\cite{quantumatk} package to construct a unit cell for multilayer heterostructures in Fig.~\ref{fig:fig1} while using experimental lattice constants of individual materials. enrolled in their database. The lattice strain  at each interface is kept below 1.5$\%$. In order to determine the interlayer distance we carry out DFT calculations with the generalized gradient approximation (GGA) for the XC functional in the Perdew, Burke and Ernzerhof (PBE)~\cite{Perdew1996} parametrization, as implemented in QuantumATK package. The Hamiltonian $\bold{H}_\mathrm{KS}$ of central region of heterostructures in Fig.~\ref{fig:fig1} and the self-energies~\cite{Rungger2008} of their semi-infinite leads are extracted from  QuantumATK calculations where we employ PBE-GGA for the XC functional; fully relativistic norm-conserving SG15-SO~\cite{Schlipf2015}  pseudo-potentials for describing electron-core interactions; and SG15 ``medium''~\cite{Schlipf2015} basis set of localized orbitals. Periodic boundary conditions are employed in the plane perpendicular to the transport direction (the $x$-direction), with 9$\times$9 $k$-point grid for the self-consistent calculation. The energy mesh cutoff for the real-space grid is chosen as 100 Hartree. 

{\em Charge conserving solution of Floquet-NEGF equations for pumped currents}.---The time-dependent NEGF formalism~\cite{Stefanucci2013,Gaury2014}  operates with two fundamental quantities~\cite{Stefanucci2013}---the retarded $\mathbf{G}^r(t,t')$ and the lesser $\mathbf{G}^<(t,t')$ Green functions (GF)---which describe the density of available quantum states and how electrons occupy those states in nonequilibrium, respectively. They depend on two times, but solutions can be sought in other representations, such as double-time-Fourier-transformed~\cite{Mahfouzi2012} GFs, ${\bf G}^{r,<}(E,E')$. In the case of periodic time-dependent Hamiltonian, they must take the form ${\bf G}^{r,<}(E,E')={\bf G}^{r,<}(E,E+n\omega)={\bf G}^{r,<}_n(E)$, in accord with the Floquet theorem~\cite{Shirley1965,Sambe1973}. The coupling of energies $E$ and $E+ n\omega$ ($n$ is integer) indicates how `multiphoton' exchange processes contribute toward the pumped current. In the absence of many-body (electron-electron or electron-boson)  interactions, currents can be expressed using solely the Floquet-retarded-GF
\begin{equation}\label{eq:floquet_GF}
[(E + \check{\bold{\Omega}})\check{\bold{S}} - \check{\bold{H}}_\mathrm{F} - \check{\bold{\Sigma}}^r(E) ]\check{\bold{G}}^r(E) = \check{\bold{1}},
\end{equation}
which is composed of ${\bf G}^r_n(E)$ as its submatrices along the diagonal. Here $\bold{H}_\mathrm{F}$ is the so-called Floquet Hamiltonian~\cite{Shirley1965,Sambe1973}; $\check{\bold{\Omega}}=\mathrm{diag}[\ldots, -2\hbar\omega\bold{1}, -\hbar\omega\bold{1},0,\hbar\omega\bold{1},2\hbar\omega\bold{1},\ldots]$; and  $\check{\bold{\Sigma}}^r(E) = \mathrm{diag}[\ldots, \bold{\Sigma}^r(E-\hbar\omega),0, \bold{\Sigma}^r(E+\hbar\omega),\ldots]$ is the Floquet self-energy matrix. The submatrices, $\bold{\Sigma}^r(E) = \sum_{p=\mathrm{L,R}} \bold{\Sigma}_p^r(E)$, along the diagonal of $\check{\bold{\Sigma}}^r(E)$ and are obtained by standard  steady-state NEGF calculations~\cite{Rungger2008} to capture the presence of semi-infinite NM leads. In the case of nonorthogonal localized orbitals as the basis set  $\ket{\phi_a}$, an infinite matrix $\check{\bold{S}}$ composed of overlap submatrices $\bold{S}$ along the diagonal with matrix elements $\mathrm{S}_{ab} = \braket{\phi_a|\phi_b}$ is also required in Eq.~\eqref{eq:floquet_GF}. 

\begin{table}[t]
	\begin{center}
		\begin{tabular}{ |c|c|c| }
			\hline
			Heterostructure & $g_{\uparrow \downarrow}$ (w/o SOC)& $g_{\uparrow \downarrow}$ (SOC) \\
			\hline
			Cu/Co/Cu& 0.7721 & 0.7721  \\ \hline
			Pt/Co/Cu & 1.4142 & 0.3866 \\ \hline
			W/Co/Cu  & 0.4191 & 0.2476 \\ \hline
			Cu/1QL-Bi$_2$Se$_3$/Co/Cu & 0.4721 & 0.1716 \\ \hline
			Cu/6QL-Bi$_2$Se$_3$/Co/Cu & 0.0002 & 0.0074\\ \hline
		\end{tabular}
	\end{center}
	\caption{The effective SMC $g_{\uparrow\downarrow}^\mathrm{eff}$, extracted from Floquet-NEGF+ncDFT computed pumped spin currents $I_\mathrm{L}^{S_z}$ in Fig.~\ref{fig:fig2} via Eq.~\eqref{eq:smceffective}, in the units of $\mathrm{10^{20} m^{-2}}$.}
	\label{tab:table1}
\end{table}

The matrices labeled as $\check{\mathbf{O}}$ are representations of operators acting in the Floquet or Sambe~\cite{Sambe1973} space, $\mathcal{H} =  \mathcal{H}_T \otimes \mathcal{H}_e$, where $\mathcal{H}_e$ is the Hilbert space of electronic states spanned by $|\phi_a\rangle$ and $\mathcal{H}_T$ is the Hilbert space of periodic functions with period $T=2\pi/\omega$ spanned by orthonormal Fourier vectors $\langle t|n \rangle = \exp(i n \omega t)$ with $n$ being integer. The space $\mathcal{H}$ is then spanned by separable states $|n \rangle  \otimes |\phi_a\rangle$. In this notation, $\check{\sigma}_\alpha = \bold{1}_T \otimes \hat{\sigma}_\alpha$ is the Pauli matrix in $\mathcal{H}$; $\bold{1}_T$ is the identity matrix in $\mathcal{H}_T$; and $\bold{1}$ is the identity matrix in $\mathcal{H}_e$. Other matrices employed in Eq.~\eqref{eq:floq_spj} below are $\check{\bold{\Gamma}}_p(E) = i [\check{\bold{\Sigma}}_p^r(E) - (\check{\bold{\Sigma}}_p^r(E))^\dagger]$; $\check{\bold{\Gamma}}(E) = \sum_{p=L,R} \check{\bold{\Gamma}}_p(E)$; and the Floquet-advanced-GF $\check{\bold{G}}^a(E) = [\check{\bold{G}}^r(E)]^\dagger$. 

In the adiabatic limit $\hbar\omega \ll E_\mathrm{F}$, justified by the Fermi energy $E_F \sim 1$ eV  of heterostructures in Fig.~\ref{fig:fig1} being much larger than typical \mbox{$\omega \sim 1$ GHz} precession frequency, the time-averaged (over a period $2\pi/\omega$) pumped spin currents flowing into the NM leads $p=\mathrm{L,R}$ (R-right) are given by~\cite{Mahfouzi2012}
\begin{multline}\label{eq:floq_spj}
I_{p}^{S_\alpha} = \frac{\hbar}{4N_\mathrm{max}A_\mathrm{cell}} \int_\mathrm{BZ}   d\mathbf{k}_\parallel \, \mathrm{Tr}[\check{\sigma}_\alpha\check{\bold{\Gamma}}_p\check{\bold{\Omega}}\check{\bold{G}}^r\check{\bold{\Gamma}}\check{\bold{G}}^a \\ - \check{\sigma}_\alpha\check{\bold{\Gamma}}_p\check{\bold{G}}^r\check{\bold{\Gamma}}\check{\bold{\Omega}}\check{\bold{G}}^a ].
\end{multline}
The pumped charge current is obtained from Eq.~\eqref{eq:floq_spj} by replacing $\check{\sigma}_\alpha \mapsto \bold{1}_T \otimes \hat{\sigma}_0$, where $\hat{\sigma}_0$ is the unit $2 \times 2$ matrix, and $\hbar/2 \mapsto e$.  Here all matrices depend on $\mathbf{k}_\parallel$ due to assumed periodicity of heterostructures depicted in Fig.~\ref{fig:fig1} within the $yz$-plane and absence of disorder, so that integration over the two-dimensional Brillouin zone (BZ) is performed and $A_\mathrm{cell}$ is the area of the unit cell in the transverse direction. 

\begin{figure}
	\centering
	\includegraphics[width=8.5cm]{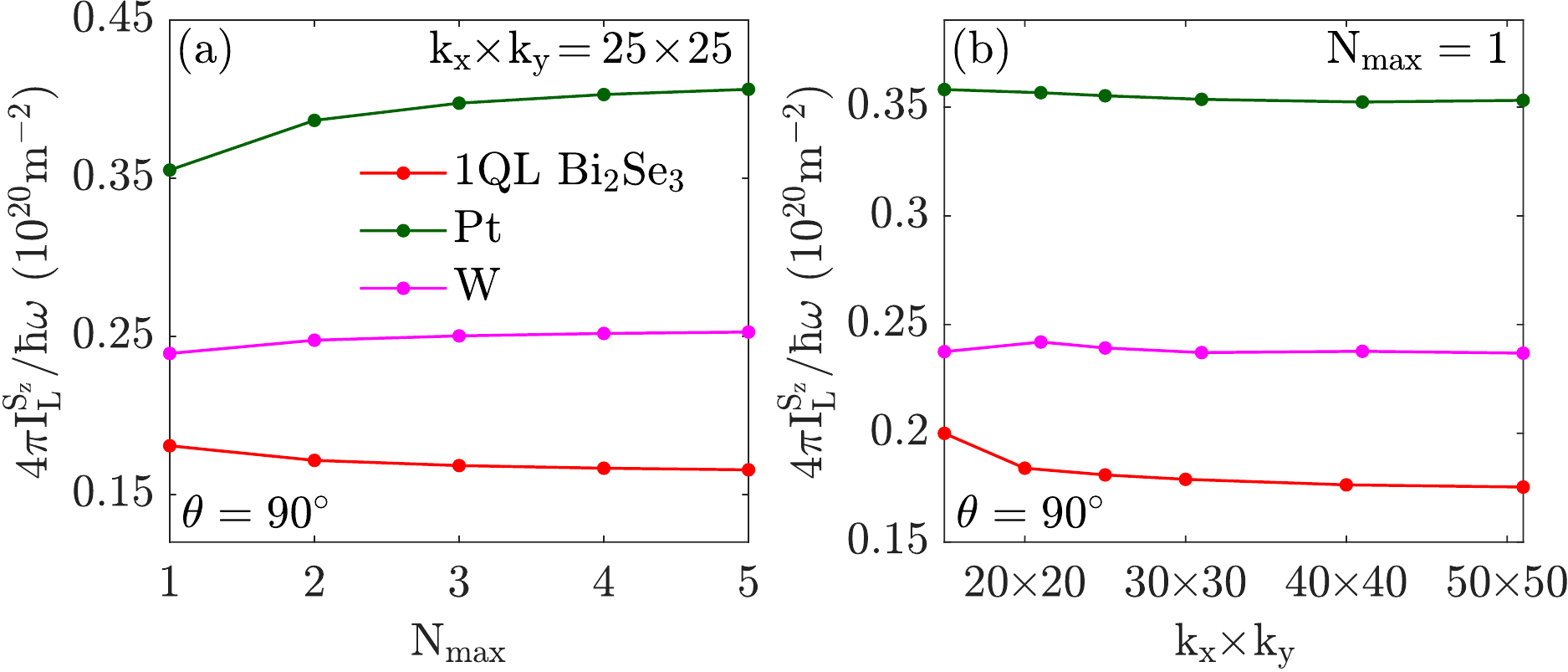}
	\caption{Convergence of DC spin current  $I_\mathrm{L}^{S_z}$ pumped into the left lead of heterostructures in Fig.~\ref{fig:fig1} with respect to: (a) 
	$N_\mathrm{max}$ determining the size of truncated Floquet Hamiltonian, at fixed grid $k_x \times k_y = 25 \times 25$ of $k$-points; (b) grid $k_x \times k_y$ at fixed  $N_\mathrm{max}=1$. The frequency is \mbox{$\hbar\omega = 1$ meV} and $\theta=90^\circ$.}
	\label{fig:fig3}
\end{figure}

Thus, the Floquet-NEGF formalism replaces the original time-dependent NEGF problem with the {\em time-independent} one delineated above, at the cost of using infinite-dimensional matrices $\check{\mathbf{O}}$ due to infinite dimensionality of $\mathcal{H}_T$. However, in practice finite $|n| \le N_\mathrm{max}$ is chosen where this range can be expanded until the answer converges, thereby yielding a {\em nonperturbative} result. Note that trace in Eq.~\eqref{eq:floq_spj}, $\mathrm{Tr} \equiv  \mathrm{Tr}_e \mathrm{Tr}_T$, is summing over contributions from different subspaces of $\mathcal{H}_T$ so that the denominator includes $2N_\mathrm{max}$ to avoid double counting. The part of the trace operating in $\mathcal{H}_T$ space ensures that at each chosen $N_\mathrm{max}$ charge current is conserved, $I_\mathrm{L} \equiv I_\mathrm{R}$, unlike some other solutions~\cite{Wang2003,Kitagawa2011} of Floquet-NEGF equations where current conservation is ensured only in the limit $N_\mathrm{max} \rightarrow \infty$. Figure~\ref{fig:fig3}(a) demonstrates  that $I^{S_z}_\mathrm{L}$ in TI/FM and HM/FM heterostructures from Fig.~\ref{fig:fig1} converges already for $N_\mathrm{max}=4$ since using  $N_\mathrm{max}=5$ leads to relative change by less than 1\%. We also confirm [Fig.~\ref{fig:fig3}(b)] that 25$\times$25 grid of $k$-points is sufficient to ensure that a relative change is less than 1$\%$ when increasing their number further. 

{\em Results for pumped spin current and effective SMC}.---To reduce the computational expense, we use $N_\mathrm{ph}=2$ and $k_x \times k_y=25 \times 25$ for angular dependence of $I_\mathrm{L}^{S_z}$ and $I_\mathrm{R}^{S_z}$ in Fig.~\ref{fig:fig2}. In order to understand the effect of SOC, Fig.~\ref{fig:fig2}(a),(b),(e),(f) shows results with SOC switched off in ncDFT calculations, while in Fig.~\ref{fig:fig2}(c),(d),(g),(h) SOC is switched on. We also consider Cu/Co/Cu  heterostructure [dotted line in Fig.~\ref{fig:fig2}(a)--(d)] as a reference system~\cite{Carva2007}. Since such heterostructure is inversion symmetric, no SOC effects can appear at its interfaces~\cite{Dolui2017}, as confirmed by identical results in Fig.~\ref{fig:fig2}(a) vs.  Fig.~\ref{fig:fig2}(c). Dotted lines in Figs.~\ref{fig:fig2}(b) and  Fig.~\ref{fig:fig2}(d) for pumped spin current into the right Cu lead are identical to those in Figs.~\ref{fig:fig2}(a) and   Fig.~\ref{fig:fig2}(c), respectively, as expected from the left-right-symmetry of Cu/Co/Cu heterostructure (thereby confirming that results are properly converged). 

Upon switching on SOC in ncDFT calculations, we find that pumped current $I_\mathrm{L}^{S_z}$ and the corresponding SMC in Table~\ref{tab:table1} are consistently reduced. An apparent exception is Cu/6QL-Bi$_2$Se$_3$/Co/Cu system, where these quantities are enhanced $\sim$ 50 times by SOC. But this is trivially explained by the fact that SOC generates  topologically protected metallic surface states, with Dirac cone energy-momentum dispersion  inside the bandgap of Bi$_2$Se$_3$. The surface states can hybridize with Co states~\cite{Marmolejo-Tejada2017} and penetrate as evanescent states into the bulk of 6QL-Bi$_2$Se$_3$ to facilitate conduction. On the other hand, 1QL of  Bi$_2$Se$_3$ remains insulating even in the presence of SOC~\cite{Zhang2010} due to minigap opening at the Dirac point. 

{\em Discussion.---}We note that calculations~\cite{Jamali2015,Chen2015} using simplistic models of FM/NM heterostructures have predicted increase of pumped spin current with SOC. However, SOC was artificially introduced in such models as strictly two-dimensional effect, thereby neglecting that even if such interfacial SOC can enhance spin pumping, the pumped spin current could subsequently experience spin memory loss~\cite{Dolui2017,Belashchenko2016} deeper inside the HM left lead or at the left Cu/TI interface. The extraction of SMC via Eq.~\eqref{eq:alpha} from first-principles calculations of Gilbert damping in HM/FM systems as a function of F layer thickness has also found increase of SMC when SOC is switched on. However, this could be due to combined effect of SOC and  hybridized FM and HM states which enhances the density of states at the Fermi level and thereby $\alpha$ as an effect unrelated to SMC~\cite{Barati2014,Barati2014}.     

{\em Conclusions.---}In conclusion, by combining the Floquet theory of periodically driven quantum systems~\cite{Shirley1965,Sambe1973} with time-dependent 
NEGFs~\cite{Mahfouzi2012,Mahfouzi2014} and ncDFT calculations, we have developed a first-principles quantum transport method which makes it possible to compute directly pumped spin current by precessing magnetization in realistic and ubiquitous in spintronics  HM/FM and TI/FM heterostructure with strong SOC effects around the interface. Such effects consistently reduce pumped spin current and the effective SMC extracted from it. We note that, in general, Floquet theorem does not allow one to transform time-dependent DFT Hamiltonian to a time-independent Floquet-DFT~\cite{Kapoor2013}. Instead, one apparently has to perform self-consistent time-dependent DFT calculations~\cite{Huebener2017} in the presence of time-periodic potential, in contrast to our first-principles Hamiltonian in Eq.~\eqref{eq:td_ham} where harmonic potential is introduced {\em a posteriori} into the converged KS Hamiltonian of static ncDFT. Nevertheless, due to small frequency of microwave radiation when compared to the Fermi energy of heterostructures in Fig.~\ref{fig:fig1}, $\hbar \omega \ll E_F$, we expect tiny perturbation of electronic density already converged by static ncDFT. In other words, our much less expensive calculations are in the spirit of linear-response theory and first-principles scattering matrix calculation~\cite{Xia2002,Zwierzycki2005,Carva2007,Liu2014a} of SMC where one also employs converged static KS Hamiltonian (but without SOC) as an input.

\begin{acknowledgments}
This work was supported by DOE Grant No. DE-SC0016380. The supercomputing time was provided by XSEDE, which is supported by NSF Grant No. ACI-1053575.
\end{acknowledgments}


\end{document}